\documentclass[a4paper,preprintnumbers,amsmath,amssymb,nofootinbib,floatfix]{revtex4}

\usepackage{color}
\usepackage{graphicx}
\usepackage{dcolumn}
\usepackage{bm}
\usepackage{latexsym}
\usepackage{epsfig}
\usepackage{rotating}
\usepackage{bm}

\newcommand{\be}{\begin{equation}}
\newcommand{\ee}{\end{equation}}
\newcommand{\beqq}{\setlength\arraycolsep{2pt}\begin{eqnarray}}
\newcommand{\eeqq}{\vspace{0cm} \end{eqnarray}}
\newcommand{\bea}{\begin{eqnarray}}
\newcommand{\eea}{\end{eqnarray}}

\begin{document}

\title{Partition function for a mass dimension one fermionic field and the dark matter halo of galaxies}

\author{S. H. Pereira}\email{shpereira@gmail.com}

\author{Richard S. Costa}

\affiliation{Universidade Estadual Paulista (Unesp)\\Faculdade de Engenharia, Guaratinguet\'a \\ Departamento de F\'isica e Qu\'imica\\ Av. Dr. Ariberto Pereira da Cunha 333\\
12516-410 -- Guaratinguet\'a, SP, Brazil
}




\begin{abstract}

This work study the finite temperature effects of a mass dimension one fermionic field, sometimes called Elko field. The equilibrium partition function was calculated by means of the imaginary time formalism and the result obtained was the same for a Dirac fermionic field, even though the Elko field does not satisfy a Dirac like equation. The high and low temperature limits were obtained, and for the last case the degeneracy pressure due to Pauli exclusion principle can be responsible for the dark matter halos around galaxies to be greater than or of the same order of the galaxy radius. Also, for a light particle of about $0.1$eV and a density of just 0.1 particle per cubic centimeter, the value of the total dark matter mass due to Elko particles is of the same order of a typical galaxy.  Such a result satisfactorily explains the dark matter as being formed just by Elko fermionic particles and also the existence of galactic halos that go beyond the observable limit. 

\end{abstract}

\maketitle

\section{Introduction}

Finite temperature effects in quantum field theory, or thermal field theories, are studied long time ago and the main results concerning relativistic degenerate gas, super-dense nuclear matter, quark-gluon phase transition, spontaneous symmetry breaking of electroweak unification and Higgs model at finite temperature have been discussed exhaustively in several textbooks \cite{bellac,kapusta,das,love}, including astrophysical and cosmological applications in white dwarf stars, neutron stars and baryogenesis after big bang.

Most of works in thermal field theory concerns the study of equilibrium thermodynamic properties of systems formed by spin-0  real or complex scalar fields (bosonic particles), spin-${1\over 2}$ fields (fermionic Dirac like particles) and spin-1 fields (gauge vector bosons). In order to extract physical thermodynamic properties of relativistic fields at equilibrium, the standard method is by means of the study of the partition function from the statistical mechanics, constructed through a path integral over the fields. Once the partition function has been evaluated, the canonical ensemble assures that the Helmholtz free energy can be obtained, from which follows the total energy density, pressure and entropy density of the system. 

As already mentioned at the beginning, finite temperature effects for scalar fields, fermionic fields and gauge vector bosons are well known in literature. Nevertheless, in which concerns the fermionic fields, the calculations are done just for fermionic Dirac like fields, that is, fermionic fields with spin-${1\over 2}$ and mass dimension-${3\over 2}$ that satisfies a Dirac like equation. Recently, a new class of mass dimension one fermionic fields has been discovered \cite{AHL1,AHL2,ahl2011a,AHL4,WT}, sometimes called Elko\footnote{From German, {\it Eigenspinoren des ladungskonjugationsoperators}, or eigenspinor of charge conjugation}. The new fermionic fields constructed from such new spinors are natural candidates to dark matter particles in the universe, once they are eigenstate of the charge conjugation operator, being neutral and coupling very weakly or even not coupling to other particles of the standard model. The signature of such new fermionic mass dimension one particle at LHC has been proposed in \cite{dias1,alves1,clee,julio2}, cosmological applications have also been recently studied in  \cite{FABBRI,BOE4,BOE6,GREDAT,BASAK,sadja,kouwn,saj,js,asf,sajf,st,sra}, a sigma model has been obtained in \cite{elkosigma}, its Casimir effect has been calculated in \cite{elkocasimir} and an one loop effective Lagrangian has been obtained in \cite{elkoeffective}.

In this work we aims to study the finite temperature effects due to Elko fermionic fields and its contribution to high and low temperature systems formed by such fields. While Elko fields satisfy just a Klein-Gordon like equation, as a scalar field, it must satisfies anti-periodic boundary conditions, as Dirac like fields. Thus the calculation of its partition function must take into account a mixture of both properties of bosonic and fermionic fields. In Section II we present a brief review of the results of finite temperature systems, in Section III we makes a explicit calculation of the partition function for Elko fields and present the main results at high temperature limit, in Section IV we calculated the degeneracy pressure in low temperature limit and its relation to dark matter halos of galaxies. We conclude in Section V. 

\section{Finite temperature field theory}

The study of different quantum fields at finite temperature is done by means of the partition function $Z$ of the system \cite{bellac,kapusta,das,love}, which is related to the Helmholtz free energy $F$ by $Z=\exp(-\beta F)$, where\footnote{We are using units where $c=\hbar=k_B=1$} $\beta \equiv 1/T$, from which follows the thermodynamic quantities as total energy $E$, pressure $P$ and entropy $S$:
\begin{equation}
E=F+TS,\hspace{0.5cm} P=-\Bigg(\frac{\partial F}{\partial V}\Bigg)_T,\hspace{0.5cm} S=-\Bigg(\frac{\partial F}{\partial T}\Bigg)_V\,.\label{EPS}
\end{equation}

The temperature is introduced by means of the imaginary time formalism, making a rotation of the real time axis to the complex one, $x_0=t\to -i\tau$, similar to go  from Minkowski to Euclidean space-time metric, with new variables $\bar{x}^\mu=(\bar{x}_0,\bar{\bf{x}})=(ix_0,\bf{x})=(\tau,\bf{x})$. Once the system is at equilibrium (not evolving in time), the equilibrium thermodynamic temperature is introduced by means of $\tau = \beta =1/T$. 

The partition function for a standard real scalar field $\phi$ is \cite{bellac,kapusta,das,love}:
\begin{equation}
Z=N(\beta)\int_{periodic}D\phi\exp\int_0^\beta d\tau \int d^3x\mathcal{L}(\phi,\bar{\partial}_\mu\phi)\,,\label{Z1}
\end{equation}
where $N(\beta)$ is just a normalisation constant ($\beta$ dependent) and $\mathcal{L}$ is the Lagrangian density of the field $\phi$ and its derivative, with $\bar{\partial}_\mu \equiv (i\partial_\tau, \partial_i)$. The functional integral over the field must be understood to be periodic in the interval $0<\tau<\beta$,
\begin{equation}
\phi(\tau=0,\bf{x})=\phi(\tau=\beta,\bf{x})\,.
\end{equation}

For a general fermionic field $\psi$ the partition function is:
\begin{equation}
Z=N(\beta)\int_{antiperiodic}D\bar{\psi}D\psi\exp\int_0^\beta d\tau \int d^3x\,\mathcal{L}(\psi,\bar{\psi})\,,\label{Z2}
\end{equation}
where $N(\beta)$ is also a normalisation constant and $\mathcal{L}$ is the Lagrangian density of the field $\psi(\bar{x})$ (and its dual $\bar{\psi}(\bar{x})$) and the functional integral over the fields must be anti-periodic in the interval $0<\tau<\beta$, 
\begin{equation}
\psi(\tau=0,\bf{x})=-\psi(\tau=\beta,\bf{x})\,.
\end{equation}

A standard real scalar field $\phi$ satisfies a Klein-Gordon like Lagrangian:
\begin{equation}
\mathcal{L}(\phi,\bar{\partial}_\mu\phi)={1\over 2}\partial^\mu\phi\partial_\mu\phi-{1\over 2}m^2\phi^2\,,
\end{equation}
while for a standard Dirac free-field, the Lagrangian is:
\begin{equation}
\mathcal{L}(\psi,\bar{\psi})=\bar{\psi}(x)(i\gamma^\mu\partial_\mu-m)\psi(x)\,.\label{lagranD}
\end{equation}
However, for a mass dimension one fermionic field $\lambda(x)$ (and its dual $\stackrel{\neg}{\lambda}$) the Lagrangian is of Klein-Gordon type:
\begin{equation}
\mathcal{L}(\lambda,\stackrel{\neg}{\lambda})={1\over 2}\partial^\mu\stackrel{\neg}{\lambda}\partial_\mu\lambda-{1\over 2}m^2\stackrel{\neg}{\lambda}\lambda\,.\label{lagran}
\end{equation}
Here is the main difference concerning the calculation of the partition function involving different kind of fields. While the calculation of (\ref{Z1}) for bosonic fields (satisfying a Klein-Gordon like equation) must be periodic in $\tau$, the calculation of (\ref{Z2}) for Dirac fields must be anti-periodic. In the case of a mass dimension one fermionic field as Elko, although satisfying a Klein-Gordon like equation, the partition function must be done with anti-periodic boundary condition, as in (\ref{Z2}). As far as we know, this calculation has not been done so far.

\section{Partition function for a mass dimension one fermionic field}

{In order to calculate the partition function for a mass dimension one fermionic field satisfying (\ref{lagran}) we must be careful. We start with the functional integral in Minkowski space-time: 
\begin{equation}
Z=N\int_{antiperiodic}D\stackrel{\neg}{\lambda}D\lambda\exp\Bigg(\frac{i}{2}\int d^4x'\int d^4x\,\stackrel{\neg}{\lambda}({x}')\mathbf{A}({x}',{x})\lambda({x})\Bigg)\,,\label{Z3}
\end{equation}
where 
\begin{equation}
\mathbf{A}({x}',{x})=-({\partial}'_\mu{\partial}^\mu+m^2) \delta({x}'-{x})\mathbf{I}\,,
\end{equation}
is the Klein-Gordon like operator that comes from (\ref{lagran}) after a partial integration and $\mathbf{I}$ stands for a $4\times 4$ identity matrix. The functional integral over Grassmannian like functions in Minkowski space-time is well established}\footnote{{According to \cite{love}, for $n$ Grassmann variables $\theta_1, ..., \theta_n$ satisfying anti commuting relations as $\{\theta_i, \theta_j\}=0$ we have $\int d\theta_i = 0$ and $\int d\theta_i \theta_i=1$. For a Gaussian integral over $n$ Grassmann variables, $I_n\equiv \int d\theta_1\,...d\theta_n\exp(-{1\over 2}\Theta^T \mathbf{A} \Theta)$, where $\mathbf{A}$ is a real anti-symmetric matrix and $\Theta$ a column vector with components $(\theta_1, ..., \theta_n)$. For $n$ odd we have $I_n=0$ and for $n$ even $I_n=(\det \mathbf{A})^{1/2}=\exp({1\over 2}Tr\ln \mathbf{A})$. For complex Grassmann variables the generalisation is $\int d\theta^{*}_1 d\theta_1\,...\int d\theta^{*}_n d\theta_n\exp(-{1\over 2}\Theta^\dagger \mathbf{B} \Theta)=\det({1\over 2}\mathbf{B})$, where $\mathbf{B}$ is a skew Hermitean matrix. For a complex Gaussian, as occurs in a Minkowski space, the generalisation is  $\int D\varphi^{*} D\varphi\exp(i\varphi^{*}\mathbf{C}\varphi)=\det(i\mathbf{C})= \exp Tr \ln(i\mathbf{C})$, where $\mathbf{C}$ is Hermitean.}}, {and the path integral over $\stackrel{\neg}{\lambda}$ and $\lambda$ can be done, giving:
\begin{equation}
Z=N\exp\big(\mathrm{Tr}\ln {i\over 2}\mathbf{A}\big)\,.
\end{equation}
Once the functional integral has been done in Minkowski space, we make a rotation to Euclidean space-time $\bar{x}^\mu$ in order to establish contact with thermodynamic by means of $\tau$. The trace must be evaluated in the Euclidean space by introducing the Fourier transform of the field as:}
\begin{equation}
\lambda(\bar{x})={1\over \beta}\sum_n\int\frac{d^3p}{(2\pi)^3}\mathrm{e}^{-i\omega_n\tau}\mathrm{e}^{i\bf{p}\cdot \bf{x}}\tilde{\lambda}(\omega_n,\bf{p})\,,
\end{equation}
with a similar one for the dual $\stackrel{\neg}{\lambda}(\bar{x})$, and where the sum is done over the Matsubara frequencies for fermion
\begin{equation}
\omega_n=\frac{(2n+1)\pi}{\beta}\,,\hspace{1cm} n \textrm{ integer}.
\end{equation}
We obtain:
\begin{equation}
\mathrm{Tr}\ln \mathbf{A}=4\int_0^\beta d\tau \int d^3x{1\over \beta}\sum_n\int\frac{d^3p}{(2\pi)^3}\ln(\omega_n^2+{\bf p}^2+m^2)\,,
\end{equation}
and performing the $\tau$ integration and the fermionic Matsubara frequency sum\footnote{See \cite{love} for further details.}, we obtain finally the partition function as:
\begin{equation}
Z=N(\beta)\exp\Bigg\{2\int d^3x \int\frac{d^3p}{(2\pi)^3}\Bigg[\beta \sqrt{{\bf p}^2+m^2}+ 2 \ln\bigg(1+\exp(-\beta \sqrt{{\bf p}^2+m^2})\bigg) + C\Bigg]\Bigg\}\,,
\end{equation}
where $C$ is a $\bf{p}$ independent constant. The Helmholtz free energy is:
\begin{equation}
F=-{1\over \beta}\ln Z = -2\int d^3x \int\frac{d^3p}{(2\pi)^3}\Bigg[\sqrt{{\bf p}^2+m^2}+ {2\over \beta} \ln\bigg(1+\exp(-\beta \sqrt{{\bf p}^2+m^2})\bigg)\Bigg]\,,\label{F1}
\end{equation}
where the constant $C$ has been cancelled against $N(\beta)$. The first term inside integral represents the divergent zero temperature contribution that must be removed by regularisation and the second term the temperature dependent one. It is interesting to notice that such result is exactly the same that one obtained by standard Dirac fermion, even though the Lagrangian for the mass dimension one Elko field is very different from the Dirac case.

Taking just the temperature dependent term of (\ref{F1}) and making an integration by parts we are left with:
\begin{equation}
{F\over V}=-\frac{2}{3\pi^2}\int_0^\infty\frac{p^4}{\sqrt{p^2+m^2}}\frac{1}{\textrm{e}^{\beta\sqrt{p^2+m^2}}+1}dp\,,\label{F2}
\end{equation}
where $V=\int d^3x$. In the last term we recognise the occupation number that characterises the Fermi-Dirac distribution function, $n_\varepsilon\equiv [\textrm{e}^{\beta(\varepsilon-\mu)}+1]^{-1}$, with relativistic energy $\varepsilon=\sqrt{p^2+m^2}$ and null chemical potential $\mu$.

In the high temperature limit $(T>>m)$ the integral in (\ref{F2}) is easily calculated \cite{bellac,kapusta,das,love} and the results for the Helmholtz free energy density, energy density, pressure and entropy density from (\ref{EPS}) are:
\begin{equation}
\frac{F}{V}=-\frac{7\pi^2T^4}{180}, \hspace{0.5cm}\frac{E}{V}=\frac{7\pi^2T^4}{60}, \hspace{0.5cm} P =\frac{7\pi^2T^4}{180}, \hspace{0.5cm}\frac{S}{V}=\frac{7\pi^2T^3}{45}.
\end{equation}

We can summarise the results including the well known results for bosonic fields, Dirac fermions and now the contribution from Elko fields, generalising the results of \cite{love}:
\begin{eqnarray}
\frac{F}{V}&=&-\frac{\pi^2T^4}{90}(N_B+{7\over 8}N_D + {7\over 8}N_E),\\
\frac{E}{V}&=&\frac{\pi^2T^4}{30}(N_B+{7\over 8}N_D + {7\over 8}N_E),\\
P&=&\frac{\pi^2T^4}{90}(N_B+{7\over 8}N_D + {7\over 8}N_E),\\
\frac{S}{V}&=&\frac{2\pi^2T^3}{45}(N_B+{7\over 8}N_D + {7\over 8}N_E),
\end{eqnarray}
where $N_B=1$ for neutral scalar field, $N_B=2$ for neutral gauge field, $N_D=4$ for a Dirac field, $N_D=2$ for a Weyl field and $N_E=4$ for a Elko field.

\section{Degeneracy pressure and dark matter halo of galaxies}

The low temperature limit is much more interesting to study and is related to the degeneracy pressure of a fermionic system. In the presence of a non-null chemical potential $\mu$, the limit $T\to 0$ makes the occupation number $n_\varepsilon$ to behave as a step function which stays constant at the value 1 for $\varepsilon<\mu_0$ and at the value 0 for $\varepsilon>\mu_0$. Thus, at $T=0$ all single-particle states are completely filled with one particle per state up to $\varepsilon=\mu_0$, following the Pauli exclusion principle, while all single-particle states with  $\varepsilon>\mu_0$ are empty. This makes the integral on (\ref{F2}) easy to be calculated and the Fermi energy is defined as $\varepsilon_F = \mu_0$, which also defines the Fermi momentum $p_F$ at which the integral must be done. For the degeneracy pressure $P_0$ at $T\to 0$ we have:
\begin{equation}
P_0=-\frac{\partial F}{\partial V}=\frac{2}{3\pi^2}\int_0^{p_F}\frac{p^4}{\sqrt{p^2+m^2}}dp\,.
\end{equation}
Apart from constant terms, the integration is exactly that one for a relativistic particle in classical statistical systems \cite{pathria}. The Fermi momentum can be written as $p_F=(3n/8\pi)^{1/3}$, where $n\equiv N/V$ is the particle number density\footnote{For photons in the whole universe, $n_\gamma \simeq 422$cm$^{-3}$ and for neutrinos $n_\nu \simeq 115$cm$^{-3}$ at present time \cite{kolb}.}. Defining $x\equiv p_F/m$, the integral can be done \cite{pathria} and in the limits $x<<1$ and $x>>1$ it is given by, respectively:
\begin{equation}
P_0\simeq\frac{8\pi m^4}{15}x^5\,,\hspace{2cm}P_0\simeq\frac{2\pi m^4}{3}x^4\,.\label{P0}
\end{equation}
Given the values of $n$ and $m$ the degeneracy pressure can  be calculated. 

Suppose that the whole universe was filled with a gas of free Elko particles that were at thermal equilibrium with all matter in the past\footnote{Although Elko does not couple to electromagnetic fields, at very high energies it could couple to Higgs fields, which would be responsible to thermalize the Elko particles with other matter fields.}. Along the universe evolution, the temperature decreases and today it must be very small, satisfying the condition $T\to 0$, similar to a cosmic microwave background temperature of about $2.75$K. As good candidates to dark matter, the Elko particles do not interact electromagnetically and falls within the gravitational potential wells around the galactic nuclei. But the gravitational attraction must be counterbalanced by the degeneracy pressure due to Pauli exclusion principle, reaching equilibrium within a ray $R$, similar to what happens in a white dwarf star or a neutron star before collapsing. The equilibrium equation can be written as \cite{pathria}:
\begin{equation}
P_0(R)=\frac{\alpha}{4\pi}\frac{GM^2}{R^4}\,,
\end{equation}
where $M=nmV$ is the total mass within a volume $V$, $G$ is the gravitational constant and $\alpha\simeq 1$ is a constant whose value depends upon the nature of the variation of the number density inside the gas. By supposing a density number $n\simeq 0.1$cm$^{-3}$ and a mass $m\simeq 0.1$eV, we have $x\simeq 0.00028$ and the first approximation in (\ref{P0}) is valid. In this case, the equilibrium radius $R=R_0$ can be written as:
\begin{equation}
R_0=\frac{3^{4/3}}{20\pi^{5/6}}\sqrt{\frac{5}{\alpha G}}\Bigg(\frac{1}{n^{1/6}m^{3/2}}\Bigg)\,.\label{R}
\end{equation}
It can be seen that greater the values of $m$ or $n$ lower the value of the radius, as expected.

For the above values of $m$ and $n$ we found $R_0\simeq 4.7\times 10^{26}$cm, four orders greater than the radius of our galaxy\footnote{For the Milk-Way or Andromedae we have $R\simeq 50,000$l.y. $\simeq 4.7\times 10^{22}$cm and $M\simeq 10^{12}M_\circ \simeq 2\times 10^{45}$g, where $M_\odot$ is the solar mass.}. Such very rough estimate shows that dark matter formed by Elko particles may be uniformly distributed beyond the radius of the galaxy, as indicated by several observations, maintained at equilibrium due to the degeneracy pressure of its fermionic particles. Moreover, for these values of mass $m$, density number $n$ and radius $R_0$, we found the total dark matter mass inside a sphere of radius $R_0$ as $M=nmV = 4.3\times 10^{78}$eV $=7.7\times 10^{45}$g, which is of the same order of the total mass $M$ of a typical galaxy like ours$^7$. In the limit of very low density, $n\simeq 10^{-10}$cm$^{-3}$, and high mass, $m\simeq 1.0$GeV, we find very low values of radius and total mass, namely $R_0\simeq 10^{13}$cm and $M\simeq 10^6$g, incompatible with observations. 

{An interesting and complete analysis considering the possibility that dark matter halos are described by the Fermi-Dirac distribution at finite temperature was done in \cite{chavanis}, where dark matter could be a self-gravitating quantum gas made of massive neutrinos at statistical equilibrium or if dark matter can be treated as a self-gravitating collisionless gas.}

\section{Conclusion}

The partition function for a mass dimension one fermionic field was calculated and the result is the same as for Dirac like fermions. Although these new kind of fields not obey a Dirac like equation, the finite temperature effects and thermodynamic properties obtained by the standard method of Matsubara frequencies sum for fermions are exactly the same. This opens the possibility to use very known results from thermofield dynamics to systems of Elko particles. In particular, results for energy density, entropy density and pressure at high temperatures are the same of standard Dirac particles. 

The low temperature limit exhibits the phenomenon of degeneracy pressure, which could be responsible for maintain a dark matter halo around galaxies nuclei. For a low mass Elko particle of about $m\simeq 0.1$eV and a density number of about $n\simeq 0.1$cm$^{-3}$ the equilibrium radius found is greater than the observable radius of a typical galaxy as ours, explaining the existence of a large halo of dark matter around galactic nucleus. Moreover, the total Elko mass attributed to dark matter is of the same order of the galactic masses, in good agreement to observations. Additionally, due to Elko particles do not interact electromagnetically, its distribution around the galactic nucleus is nearly spherical and uniform, modelled just by equilibrium between gravitational pressure and degeneracy pressure, while baryonic matter condensates at the centre of the galaxies, forming heavier particles that does not suffer from degeneracy pressure and interact electromagnetically, which make it to loses energy and condensate. This shows correctly that most of the mass of the galaxy may be in the form of dark matter representing about 25\% of dark matter against less than 5\% of baryonic matter of the total content of the universe, in good agreement to the $\Lambda$CDM model data.

{It is important to notice that while some estimates at LHC searches indicate a very huge mass to Elko (about $100$GeV in \cite{clee} and 1TeV in \cite{julio2}), cosmological estimates based on some observational constraints point to very small masses (about $10^{-32}$eV in \cite{sajf,st,sra}). The value estimated here is of same order of neutrino masses, also considered a good dark matter candidate. More realistic models of varying mass distribution around the galactic nuclei need to be analysed in order to confirm Elko particles as good candidate to dark matter in the universe. Also, a more complete discussion about the thermal decoupling of Elko dark matter particles and its present temperature is an interesting subject to be studied, as it is done for WIMPS \cite{ultimo}.}

\begin{acknowledgements}
SHP acknowledges CNPq - Conselho Nacional de Desenvolvimento Cient\'ifico e Tecnol\'ogico, Brazilian research agency, for financial support, No. 303583/2018-5 and 400924/2016-1. This study was financed in part by the Coordenação de Aperfeiçoamento de Pessoal de Nível Superior - Brasil (CAPES) - Finance Code 001. RSC acknowledges support from CAPES. We thank J. M. Hoff da Silva for helpful discussions.
\end{acknowledgements}



\end{document}